\documentclass[%
 reprint,
 amsmath,amssymb,
 aps,
]{revtex4-2}

\usepackage{graphicx}
\usepackage{dcolumn}
\usepackage{bm}
\usepackage[caption=false]{subfig}

\begin{document}


\title{Comment on "Identification of the Mott Insulating Charge Density Wave State in 1T-TaS$_2$"} 
\author{Diego Pasquier}
\email{pasquierdiego@gmail.com}
\affiliation{Institute of Physics, Ecole Polytechnique F\'{e}d\'{e}rale de Lausanne (EPFL), CH-1015 Lausanne, Switzerland}

%

\maketitle
\begin{figure*}[t]
        \subfloat[\label{figa}]{
  \includegraphics[width=0.66\columnwidth]{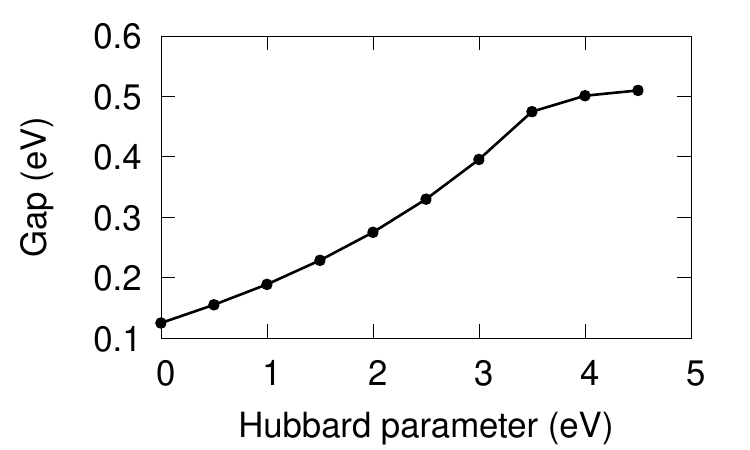}%
}\hfill
        \subfloat[\label{figb}]{
  \includegraphics[width=0.66\columnwidth]{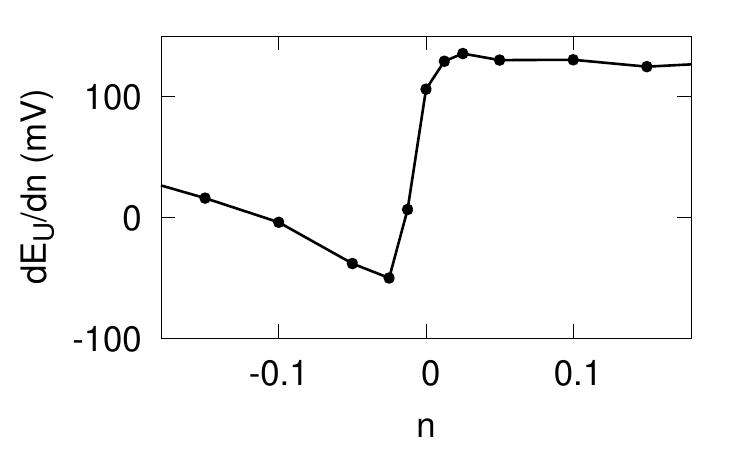}%
}\hfill
        \subfloat[\label{figc}]{
  \includegraphics[width=0.66\columnwidth]{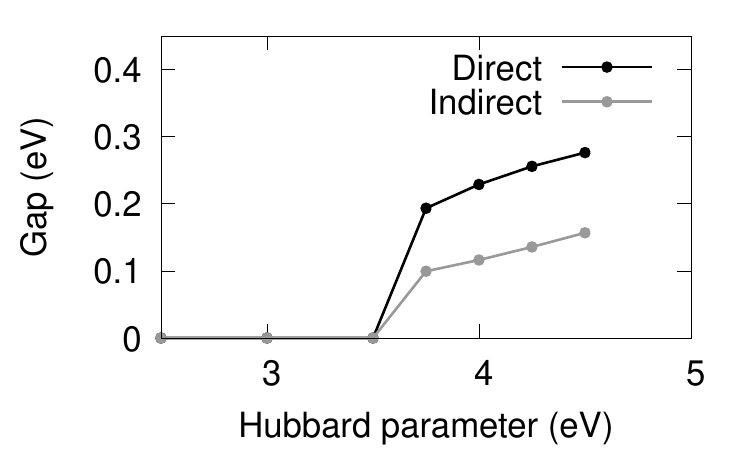}%
}
	\caption{(a) Gap at $\bar{\Gamma}$-point in monolayer TaS$_2$, calculated with DFT+$U$ (b) Derivative discontinuity of the Hubbard energy, calculated with $U=2.5$ eV. $n=0$ corresponds to charge neutrality. (c) Direct (at $\bar{A}$) and indirect ($\bar{A}\bar{\Gamma}$) gap in bulk TaS$_2$ with direct stacking, calculated with DFT+$U$ using experimental lattice parameters.}\label{fig:a}
\end{figure*}

Shin \textit{et al.} recently studied the commensurate charge-density-wave (CCDW) state of $1T$-TaS$_2$ using a generalized Hubbard DFT approach (DFT+$GOU$) that employs generalized projectors, arguing that standard DFT+$U$ with atomic-like projectors could seriously misevaluate correlations in the extended-orbital CDW state \cite{shin2021identification}.
The DFT+$GOU$ approach was shown to give $U$-dependent gap in monolayer form and predict a layer-by-layer antiferromagnetic (AFM) insulating state for bulk form.
The improved description was rationalized in terms of the revival of the derivative discontinuity, and it was concluded that the newer method resolves a decades-old inconsistency.

Given the good agreement between DFT+$U$ and experimental spectra in analogous systems \cite{chen2020strong, liu2021monolayer}, these conclusions appear very surprising.
This comment points out that the striking difference between the methods for TaS$_2$ was largely due to the use of unrealistic parameters in DFT+$U$, and that there was no compelling evidence for the conceptual superiority of DFT+$GOU$ in this system.
It is also pointed out why this Mott state can not be the ground state of bulk TaS$_2$.

For the comparison between the methods to be valid, the Hubbard parameters need to be properly renormalized when the projectors are changed. 
Parameters of few tenths of eV are order of magnitude too small for atomic projectors in related compounds \cite{darancet_three-dimensional_2014, pasquier2018charge}.
Estimating $U$ in the high-symmetry structure from DFPT (using the implementation of Ref.~\cite{giannozzi2009quantum}) gave $U=2.65$ eV for Ta, or $U=3.13$ eV for Ta and $U=7.77$ eV for S for larger correlated subspace.

Fig. 1 shows that with more realistic parameters, DFT+$U$ gives a $U$ dependence of the gap very similar to the DFT+$GOU$ results of Shin \textit{et al.}, unlike what was claimed in Ref.~\cite{shin2021identification}.  
Fig.~\ref{figb} also shows a derivative discontinuity of the Hubbard energy at half-filling of the correlated state, which is the desired feature to calculate Mott gaps as reminded by Shin \textit{et al.}
Remaining small numerical differences are expected, due to e.g. the effect of truncating intersite correlations, and it needs to be established which scheme gives best overall agreement with experiments for TaS$_2$ and the similar materials TaSe$_2$ and NbSe$_2$, that adopt the same CDW star-of-David structure.
This could be done for instance by directly comparing the results of the two methods to the spectra from Ref.~\cite{chen2020strong} for monolayer $1T$-TaSe$_2$.

Using the experimental lattice parameters from Ref.~\cite{givens1977thermal}, the AFM insulating solution for bulk form with aligned layers is more stable than the nonpolarized metallic one only for $U$ larger than $U_c \sim 3.5~$eV.
Using relaxed in-plane parameters gave narrower bandwidth ($\sim 0.35~$eV versus $0.45~$eV) and smaller $U_c \sim 3~$eV.
This sensitivity to the structure likely explains the discrepencies between previous reports using DFT+$U$: Refs.~\cite{yu2017electronic, boix2021out} report a Mott gap, while Ref.~\cite{darancet_three-dimensional_2014} does not.

The Mott state with AFM order can not be the ground state of TaS$_2$, due to several discrepancies with experiments.
First, the insulating behaviour in TaS$_2$ is only observed at the onset of a peculiar stacking order with bilayer dimerization \cite{ishiguro1991electron}, which was ignored here.
When this order collapses and the direct A stacking is realized, metallic behaviour is observed, which is consistent with the results of Fig.\ref{figc} \cite{stahl2020collapse, martino2020preferential}.
Second, the size of the energy gap inferred from transport measurment is only a few meV \cite{ngankeu2017quasi}, i.e. much smaller than what is calculated here assuming Mott physics.
Finally, bulk TaS$_2$ is not antiferromagnetic and shows no intrisic local moments formation \cite{disalvo1980paramagnetic}. 
It has recently been established that the bulk electronic structure is band rather than Mott insulating, although surface states can be correlated insulators depending on the termination type \cite{lee2021distinguishing}.

\vskip 0.5cm
\noindent
\bibliographystyle{apsrev4-1}
\bibliography{comment.bib}
\end{document}